\documentclass[12pt]{article}
\usepackage[T2A,OT1]{fontenc}
\usepackage[cp1251]{inputenc}
\usepackage{babel}
\usepackage{graphicx}

\usepackage{geometry}
{\makeatletter\@addtoreset{equation}{section}\makeatother}

\usepackage{hyperref}
\hypersetup{linkcolor=blue, citecolor=blue, colorlinks=true}

\usepackage{amsthm}
\usepackage{amssymb}
\usepackage{amsfonts, amsmath, amscd, amssymb,amsthm}
\usepackage{amsmath}
\usepackage{xcolor}

\newtheorem{theorem}{Theorem}

\newcommand{\Bibitem}[1]{\vspace*{-3.5mm}\bibitem{#1}\mbox{ }$\!\!\!\!$}

\newcommand{\Label}[1]{\label{#1}}

\newcommand{\BBox}{\rule{8pt}{8pt}}

\newcommand{\ZZ}{{Z\!\!\! Z}}
\newcommand{\RR}{{I\!\! R}}

\newcommand{\zd}{{{\ZZ}^{d}}}

\newcommand{\dis}[1]{{\displaystyle{#1}}}

\newcommand{\ssum}{\mathop {\sum}}

\newcommand{\ssup}{\mathop {\sup}}
\newcommand{\ttimes}{\mathop {\times}}

\newcommand{\iinf}{\mathop {\inf}}
\newcommand{\llim}{\mathop {\lim}}

\newcommand{\dd}{\partial}

\newcommand{\vv}{\vert}
\newcommand{\vvv}{\Vert}

\newcommand{\w}[1]{{\widetilde{#1}}}
\newcommand{\ffrac}[2]{ {\displaystyle{ \frac{#1}{#2} }}}
\newcommand{\ga}{\gamma}

\newcommand{\al}{\alpha}
\newcommand{\de}{\delta}
\newcommand{\la}{\lambda}
\newcommand{\vph}{\varphi}

\newcommand{\pr}{\prime}
\newcommand{\all}{{\forall}\, }
\newcommand{\is}{{\exists}\, }
\newcommand{\vep}{\varepsilon}
\newcommand{\ee}{{\dis{e}}}

\newcommand{\think}[3]{
\vspace{0.3cm}
\par\noindent{\bf #1}\ {\bf #2}  \ {\it #3}
\vspace{0.3cm}
}

\newcommand{\gml}{{{\cal G}\{ \mu_\Lambda\} }}

\title{Decay of correlations and uniqueness of Gibbs
lattice systems with non-quadratic interaction}
\author{\textbf{A. Val. Antoniouk and A.Vict. Antoniouk}\\
\footnotesize Institute of Mathematics NAS of Ukraine\\
\footnotesize Tereshchenkivska 3, Kiev, 01 024 Ukraine,\\
\footnotesize E-mail: antoniouk.a [at] gmail.com}
\date{}
\begin{document}

\maketitle

\begin{abstract}
We aim this paper to develop the classical lattice models with unbounded
spin to the case of non-quadratic polynomial interaction.
We demonstrate that the distinct relation between the
growths of potentials leads to the uniqueness and the fast decay of
correlations for Gibbs measure.\footnote{Journal of Mathematical Physics, {\bf 37}, N$^o$ 11 (1996).}
\end{abstract}

There is an approach initiated in the papers \cite{Do1,Do2,LaRu} to the
description of the probability measures on infinite dimensional spaces
in the terms of conditional distributions. This approach has already
found its non-trivial applications to the natural construction of the
different models in the Quantum Field Theory, Mathematical and
Statistical Physics \cite{GlJa,MaMi,SiPf,Sin}.

There were obtained the effective
criteria on the existence and  uniqueness  of such systems, see
Dobrushin's criterion \cite{Do1,Do2,Do3}, Dobrushin--Shlosman mixing
condition \cite{DoSh1,DoSh2}.
In the essence of the
Dobrushin's type criteria lie the keen variational estimates on the
one-point conditional measures, which admit iteration and application of the fixed
point arguments. Moreover, such estimates were used in the applications
to the lattice spin systems of the statistical physics
to the study of decay of correlations, differentiability of pressure
and the connected questions
\cite{COPP,Fo,GrD,GrP,Kl,Ku,Sin}.

In the noncompact spin case the check of Dobrushin's conditions is
rather complicated by principal unboundedness of interaction potentials.
The results in this direction were mainly
centered around the regular interactions
\cite{BePi,COPP,Is,LePr,Pic,Roy,Ru}, i.e.
when the many-point potentials in the Hamiltonian
\newpage
\noindent admit the quadratic domination, for example with the quadratic two-point potentials
\vspace*{-1mm}\[
H(x)=\ssum\limits_{k\in\zd}F(x_k)+\la\ssum\limits_{k,j\in\zd}
b_{k-j}(x_k-x_j)^2\vspace*{-1mm}\]

On the other hand, a wide class of models with nonregular interaction,
associated with
massless free lattice field, perturbed by $(\nabla\varphi )^4$
\vspace*{-1mm}\[
H(x)=\ssum\limits_{\vv \, k-j\,\vv =1}(x_k-x_j)^2
+\la\ssum\limits_{\vv\, k-j\,\vv =1}(x_k-x_j)^4,
\vspace*{-1mm}\]
has already obtained a detail investigation through various
techniques \cite{BFLS,BrYa,GaKu,MaSe,NaSp}.
In particular, it was shown
 that the exponential decay of
correlations for such systems does not occur for all $\la >0$
\cite{BFLS}, i.e. the Dobrushin uniqueness
technique does not work for such Hamiltonians.

In this paper we demonstrate that there is a wide class of the Gibbs
lattice systems, which {\it do not fulfill}
the regularity assumption but
have the fast decay of correlations.
We aim this paper to show that the application of the Dobrushin's
uniqueness technique for the Hamiltonian
\vspace*{-1mm}\[H(x)=\ssum\limits_{k\in\zd}F(x_k)+
\la\ssum\limits_{k,j\in\zd}G_{k-j}(x_k-x_j)\vspace*{-1mm}\]
with {\it polynomials} $\{ G_j\}$ requires the distinct
correlations between the growths of the interaction
potentials $\{ G_j\}$ and selfaction $\{
F\}$. This gives us possibility to treat the problem on the
existence, uniqueness and the {\it exponentially} fast decay of correlations
in the case of {\it non-quadratic} polynomial interaction. We base our
investigation on the scheme of papers \cite{Do1,Do2,Do3,Fo,GrD,Ku}
and apply Brascamp-Lieb inequality \cite{BrLi} to obtain
estimates on the distance in variations.

\vspace{6mm}

Consider $\zd$ to be $d$- dimensional integer lattice, to each point of
which corresponds the linear spin space $\RR^1$. Let $\gml$ denote the
set of Gibbs measures \cite{Do1,Do2,LaRu} on the product $\sigma$-algebra
on $\RR^\zd$. It means that the corresponding
conditional measures $\{\mu_\Lambda\}$ in
the finite volumes of the lattice $\Lambda\subset\zd$ are defined by
\vspace*{-2mm}\begin{equation}
d\mu_\Lambda =\ffrac{1}{Z_\Lambda}\exp \{ -\la \ssum\limits_{\{
k,j\}\cup\Lambda \neq
\emptyset}G_{k-j}(x_k-x_j)\}\ttimes\limits_{k\in\Lambda}
\ee^{-F(x_k)}dx_k
\Label{1}\vspace*{-2mm}\end{equation}
i.e. for all cylinder bounded functions $f\in C_{b,cyl}(\RR^\zd )$ we
have $\mu (\mu_\Lambda (f))=\mu (f)$, where $\mu (f)$ denotes the
expectation and $Z_\Lambda$ is a normalization factor.

We put the following conditions on the interactive potentials $\{
F,G_j\}$ in the Gibbs measure (\ref{1}).
\newpage
\begin{description}
\item[A.] \ \ Self-action potentials $F\in C^2(\RR^1)$, fulfill $F(0)=0$,\
$\is \vep >0\ \ \iinf\limits_{x\in\RR}F^{\pr\pr}(x)\geq\vep$ and
have no more than the exponential growth on the infinity $\is
c,a$: $\all x\ \vv F(x)\vv\leq c\ee^{a\vv x\vv}$;

\item[B.] \ \ Interaction potentials $\{ G_j\in
C^2(\RR^1)\}_{j\in\zd\backslash \{ 0\}}$,
fulfill $G_j(0)=0$, $\all j\in\zd\backslash \{ 0\}\ \all x\in \RR^1:\
\  G_j^{\pr\pr}(x)\geq 0$ and $\is r_0\;
\all j: \ \vv j\vv >r_0\ \Rightarrow  \ G_j\equiv 0$

\item[C.] \ \ Growth condition $\all k\in\zd\ \vv k\vv\leq r_0$ \ \
$\ssup\limits_{x_k,x_0\in\RR^1}\ffrac{
\vv G^{\pr\pr}_k(x_k-x_0)\vv}{\sqrt{F^{\pr\pr}(x_k)}
\sqrt{F^{\pr\pr}(x_0)}}<\infty$
\end{description}

Immediately remark that the condition C states the
domination of the one-point potentials over the interaction.
It always holds for the quadratic and less than quadratic interaction
due to $\sup \vv G^{\pr\pr}\vv\leq const$.
Actually condition C permits to consider
the interaction  $\{ G_j\}$ to be of polynomial type.

\vspace{2mm}

The following theorem states the uniqueness and the
exponentially fast decay of correlations for the Gibbs measure (\ref{1}).
The existence of such measure and finiteness of its moments is shown
in Theorem 2.

\begin{theorem}\Label{t1}\ 
Suppose conditions A-C hold and the set of measures $\mu\in\gml$, which satisfy
\begin{equation}
m_\mu =\ssup\limits_{k\in\zd}\int\limits_{\RR^\zd}\rho^2(x_k,0)
d\mu <\infty,\ \ \ \rho (x,y)=\int^x_y\sqrt{F^{\pr\pr}(s)}\, ds,
\Label{f4}\end{equation}
is nonempty. Denote
$\ga_d=\ssum\limits_{k\in\zd}\ee^{d(k,0)}\ssup\limits_{x_k,x_0\in\RR^1}
\ffrac{\vv G^{\pr\pr}(x_k-x_0)\vv}{\sqrt{F^{\pr\pr}(x_k)}\sqrt{F^{\pr\pr}(x_0)}}$
for some transitional invariant semimetric $d(k,j)$ on the lattice $\zd$.

Then $\all \la\in [0,1/\ga_d )$ measure $\w{\mu}\in\gml ,\
m_{\w{\mu}}<\infty$ is unique and has exponentially fast decay of
correlations, i.e.
\begin{equation}
\ssum\limits_{k\in\zd}\ee^{d(k,0)}\vv cov_{\w{\mu}}(f,\tau_k
g)\vv
\leq\ffrac{1}{1-\la\ga_d}(\ssum\limits_{k\in\zd}\ee^{d(k,0)}\de_k(f))
(\ssum\limits_{j\in\zd}\ee^{d(j,0)}\de_j(g))
\Label{3}\end{equation}
Above $\tau_k$ is a shift operator on vector $k\in\zd$,
\begin{equation}
\de_k(f)=\ssup\limits_{x\in\RR^\zd}\vv\ffrac{\dd_k f(x)}{
\sqrt{F^{\pr\pr}(x_k)}}\vv ,\ \ \ \ \ \
\dd_k f(x)=\ffrac{\dd f(x)}{\dd x_k},\ \ x=\{ x_k\}_{k\in\zd}
\Label{kk1}\end{equation}
Inequality (\ref{3}) is understood on the cylinder bounded
differentiable functions $f,g\in
C^1_{b,cyl}(\RR^\zd )$ such that $\ssum\limits_{j\in\zd}\ee^{d(j,0)}
\de_j(f)
<\infty$.
\end{theorem}

\noindent{\it Proof.}\ \ \ We discuss the main tool, which enables us to deal with
the polynomial interaction in the Gibbs measure.
First note that the usual estimate on the covariance \cite{BaEm,DaSi}
\begin{equation}
 cov_\mu (f,f)\equiv\int\limits_{\RR^1} (f-\int\limits_{\RR^1}f\, d\mu
 )^2 d\mu \leq\frac{1}{\vep}\int_{\RR^1}\Big\vv\ffrac{\dd f}{\dd
x}\Big\vv^2 d\mu
\Label{6}\end{equation}
for the probability measure $\mu$, $d\mu =\ee^{-F(x)}dx$ on the line
$\RR^1$, holds for {\it arbitrary} function $F\in C^2(\RR )$ such that
$F^{\pr\pr}(x)\geq \vep >0$ for all $x\in\RR^1$. Actually the above
inequality (\ref{6}) is not optimal and in the paper \cite[Th.4.1]{BrLi}
it was found that the next weighted generalization is true
\vspace*{-2mm}\begin{equation}
cov_\mu
(f,f)\leq\int_{\RR^1}\ffrac{1}{{F^{\pr\pr}(x)}}\Big\vv
\ffrac{\dd f}{\dd x}\Big\vv^2 d\mu
\Label{7}\end{equation}
with the weight $1/F^{\pr\pr}$, which in the cases when $F^{\pr\pr}$
grows on the infinity improves inequality
(\ref{6}).

Introduce the family of one-point conditional
measures $\{\mu_k\}_{k\in\zd}$
\vspace*{-2mm}\begin{equation}
d\mu_k =\frac{1}{Z_k}\exp \{-\la\ssum\limits_{j:\, j\neq
k}G_{k-j}(x_k-x_j)\} \ee^{-F(x_k)}dx_k
\Label{f1}\vspace*{-2mm}\end{equation}
where $Z_k$ is a normalization factor.
Below we also understand the measure $\mu_k$ as the operator of
conditional expectation
\vspace*{-1mm}\[\mu_k :\ C^1_{b,cyl}(\RR^\zd )\ni f\longrightarrow \mu_k(f)
\stackrel{def}{\equiv}\int_{\RR^1_k}f\, d\mu_k\in C^1_{b,cyl}(\RR^\zd
)\vspace*{-1mm}\]

The next identity for $j,k\in\zd ,\ j\neq k$
\vspace*{-1mm}\[\dd_j\mu_k(f)=\mu (\dd_j f)-\la
cov_{\mu_k}(f, \dd_j G_{k-j}(x_k-x_j))\vspace*{-1mm}\]
leads to
\vspace*{-1mm}\[\de_j (\mu_k(f))=\sup \vv\frac{\dd_j (\mu_k(f))}{
\sqrt{F^{\pr\pr}}}\vv =\vspace*{-1mm}\]
\vspace*{-2mm}\begin{equation}
=\sup \vv\mu_k (\frac{\dd_j f}{\sqrt{F^{\pr\pr}(x_j)}})-
\la cov_{\mu_k}(f,\frac{\dd_j G_{k-j}(x_k-x_j)}{
\sqrt{F^{\pr\pr}(x_j)}})\vv\leq
\Label{8}\vspace*{-2mm}\end{equation}
\vspace*{-1mm}\[\leq\de_j(f) +\la\sup \vv cov_{\mu_k}(f,\frac{\dd_j
G_{k-j}(x_k-x_j)}{\sqrt{F^{\pr\pr}(x_j)}} )\vv\vspace*{-1mm}\]

Using the convexness of $G_j$
we obtain the following consequence of the weighted
inequality (\ref{7})
\vspace*{-1mm}\[cov_{\mu_k}(f,f)\leq\int_{\RR^1}\frac{\vv\dd_k f\vv^2}{F^{\pr\pr}(x_k)+
\ssum\limits_{j\neq k}G^{\pr\pr}_{k-j}(x_k-x_j)}
d\mu_k\leq\vspace*{-1mm}\]
\vspace*{-2mm}\begin{equation}
\leq \int_{\RR^1}\frac{\vv\dd_k f\vv^2 }{{F^{\pr\pr}(x_k)}}
 d\mu_k\leq [\de_k(f)]^2
 \Label{kkk}\vspace*{-2mm}\end{equation}

Inequality (\ref{kkk}) enables us to estimate the second term in
(\ref{8})
\[\sup \vv cov_{\mu_k}(f,\frac{\dd_j
G_{k-j}(x_k-x_j)}{\sqrt{F^{\pr\pr}(x_j)}}
)\vv\leq\]
\[\leq \sup cov_{\mu_k}^{1/2}(f,f)cov_{\mu_k}^{1/2}(
\frac{\dd_j G_{k-j}(x_k-x_j)}{\sqrt{F^{\pr\pr}(x_j)}} ,
\frac{\dd_j G_{k-j}(x_k-x_j)}{\sqrt{F^{\pr\pr}(x_j)}}
)\leq\]
\[\leq \de_k(f)(\int_{\RR^1}\frac{\vv \dd_k\dd_j
G(x_k-x_j)\vv^2}{{F^{\pr\pr}(x_k)}{F^{\pr\pr}(x_j)}}
d\mu_k )^{1/2}\leq \de_k(f)\sup\frac{\vv G^{\pr\pr}_{k-j}(x_k-x_j)\vv}{
\sqrt{F^{\pr\pr}(x_k)}\sqrt{F^{\pr\pr}(x_j)}}\]

Finally from  (\ref{8}) we obtain  that
\begin{equation}
\de_j(\mu_k (f))\leq\de_j(f)+\la C_{kj}\de_k(f)
\Label{9}\end{equation}
with
\[C_{kj}=\ssup\limits_{x_k,x_j\in \RR^1}\frac{\vv G^{\pr\pr}_{k-j}(x_k-x_j)\vv}{
\sqrt{F^{\pr\pr}(x_k)}\sqrt{F^{\pr\pr}(x_j)}}\]

The estimate (\ref{9}) is a key point of the Dobrushin's uniqueness
technique and the special structure of the {\it covariance}
matrix $C_{kj}$ permits the polynomiality of $\{ G_j\}$ in
the interaction.

\vspace{3mm}
Below we follow scheme of \cite{Fo,GrD,Ku}.
The principal modification lies in the use of weighted
inequality (\ref{7}) and weighted estimate on covariances (\ref{9}).

\vspace{3mm}

\noindent 1.\ \ \ {\it Uniqueness of the Gibbs measure.} \
Like in \cite{Fo} we say that
the vector $\{ a_j\}_{j\in\zd}$ is an estimate for probability
measures $\mu ,\nu$ if $\all f\in C^\infty _{b,cyl}(\RR^\zd )$:
$\ssum\limits_{k\in\zd}\de_k(f) <\infty$ we have
\begin{equation}
\Big\vv \int_{\RR^\zd} f\, d\mu -\int_{\RR^\zd} f\,
d\nu\Big\vv\leq\ssum\limits_{j\in\zd}a_j\de_j(f)
\Label{f3}\end{equation}

For any two measures $\mu_1,\mu_2\in\gml$ with property (\ref{f4}) there
is an estimate $\w{a}=\{ \w{a}_j\equiv m_0\equiv const\}_{j\in\zd}$ with
$m_0=m^{1/2}_{\mu_1}+m^{1/2}_{\mu_2}$.
To show this, note first that for $ f\in C^1_{b,cyl}(\RR^\zd )$ with
$\ssum\limits_{k\in\zd}\de_k(f)<\infty$ we have
\[\vv f(x)-f(y)\vv\leq\ssum\limits_{i\in\zd}\de_i(f)\rho (x_i,y_i)\]
and therefore
\[\vv \int_{\RR^\zd} f\, d\mu_1 -\int_{\RR^\zd} f\, d\mu_2 \vv
=\vv\int_{\RR^\zd}(f(x)-f(0))d\mu_1
-\int_{\RR^\zd}(f(x)-f(0))d\mu_2\vv\leq\]
\begin{equation}
\leq \ssum\limits_{k\in
\zd}\de_k(f)\int_{\RR^\zd}\rho (x_k,0)\{ d\mu_1 (x)+d\mu_2 (x)\}\leq
m_0\ssum\limits_{k\in\zd}\de_k(f)
\Label{12}\end{equation}

By (\ref{9}) the operator $f\to \mu_k(f)$ preserves the
class of functions $\{ f\in C^1_{b,cyl}(\RR^\zd ):\
\ssum\limits_{k\in\zd}\de_k(f) <\infty \}$. From (\ref{9}) and
(\ref{12}) we have
\vspace*{-1mm}\[\vv \mu_1(f)-\mu_2(f)\vv =\vv (\mu_1-\mu_2)(\mu_k(f))\vv\leq\vspace*{-1mm}\]
\vspace*{-2mm}\begin{equation}
\leq \ssum\limits_{j\in\zd} \w{a}_j\de_j
(\mu_k(f))\leq\ssum\limits_{j:\, j\neq k}\w{a}_j\de_j(f)
+\la\de_k(f)\ssum\limits_{i:\, i\neq k}\w{a}_i C_{ki}
\Label{f2}\vspace*{-2mm}\end{equation}
Iterating the above estimate by choosing some enumeration
$k_1,...,k_n,..$ of the points of lattice $\zd$ one can in a purely
algebraic way achieve the following estimate, see \cite[Lemma 2.3]{Fo}
\vspace*{-1mm}\[\vv\mu_1(f)-\mu_2(f)\vv\leq\la\ssum\limits_{k\in\zd}\de_k(f)
(\ssum\limits_{j\in\zd}\w{a}_j C_{kj})\vspace*{-1mm}\]
which gives
\vspace*{-1mm}\[\vv\mu_1(f)-\mu_2(f)\vv\leq\ssum\limits_{k\in\zd}(\w{a}(\la
C)^n)_k\de_k(f)\vspace*{-1mm}\]
for all $n\geq 0$.

Due to
\vspace*{-1mm}\[\vvv \w{a}(\la C)^n\vvv_{\ell_\infty (\zd
)}=m_0\ssup\limits_{k\in\zd}\ssum\limits_{j\in\zd}\{ (\la C)^n\}_{kj} =\vspace*{-1mm}\]
\vspace*{-2mm}\begin{equation}
=m_0\ssup\limits_{k\in\zd}\la^n\ssum\limits_{j(1)\in\zd}...
\ssum\limits_{j(n-1)\in\zd
}\ssum\limits_{j\in\zd}C_{kj(1)}...C_{j(n-1)j}\leq
\Label{12a}\vspace*{-2mm}\end{equation}
\vspace*{-1mm}\[\leq m_0(\ssup\limits_{k\in\zd}\la\ssum\limits_{j\in\zd}C_{kj})^n
\leq m_0(\la \ga_d)^n\to 0,\ \ \ n\to\infty,\vspace*{-1mm}\]
we obtain the uniqueness of the Gibbs measure.

\vspace{3mm}

\noindent 2.\ \ \ {\it Decay of correlations.} \
 Fix function $g\in C^1_{b,cyl}(\RR^\zd )$
such that $\int_{\RR^\zd} g\, d\mu =1$,
$g>0$ and $\ssum\limits_{k\in\zd}\ee^{d(k,0)}\de_k(g)<\infty$.
Then measure $d\nu =g\, d\mu$ for the unique measure $\mu\in\gml$ with
property (\ref{f4}) has the same property
\[\ssup\limits_{k\in\zd}\int\limits_{\RR^\zd}\rho (x_k,0)d\nu (x)\leq
\vvv g\vvv_{C_b}m_\mu^{1/2}<\infty\]
In analog to (\ref{12}) this gives the estimate $\w{a}=\{ \w{a}_j\equiv
m_\mu^{1/2}(\vvv g\vvv_{C_b}+1)\}_{j\in\zd}$ on measures
$\mu$ and $\nu$
\vspace*{-1mm}\[\vv \mu (f)-\nu (f)\vv\leq \ssum\limits_{k\in\zd}\w{a}_k \de_k(f)
=m_\mu^{1/2}(\vvv
g\vvv_{C_b}+1)\ssum\limits_{k\in\zd}\de_k(f)\vspace*{-1mm}\]
Now we prove that if $\{ a_j\}_{j\in\zd}$ is an estimate, then $\{
\ssum\limits_{j\in\zd}a_j C_{jk}+b_k\}_{k\in\zd}$ for $b_k=\de_k(g)$
is an estimate too. Indeed
\vspace*{-1mm}\[\vv\mu (f)-\nu (f)\vv\leq\vv (\mu -\nu )_y\{ \int_{\RR_k}f(\cdot\vv y)
d\mu_k(\cdot\vv y)\}\vv +\vspace*{-1mm}\]
\vspace*{-1mm}\[+\vv\nu_y\{ \int_{\RR_k}f(\cdot\vv y)d\mu_k(\cdot\vv
y)-\int_{\RR_k}f(\cdot\vv y)d\nu_k(\cdot\vv y)\}\vv =\vspace*{-1mm}\]
\vspace*{-2mm}\begin{equation}
=\ssum\limits_{j\in\zd} \w{a}_j\de_j (\mu_k(f))
+\vv\nu_y\{ \int_{\RR_k}f(\cdot\vv y)d\mu_k(\cdot\vv
y)-\int_{\RR_k}f(\cdot\vv y)d\nu_k(\cdot\vv y)\}\vv
\Label{10}\vspace*{-2mm}\end{equation}

Using (\ref{9}) the first term in (\ref{10}) can be estimated by
\vspace*{-1mm}\[ \ssum\limits_{j\neq k}\w{a}_j \de_j (f)+\la \,\de_k(f)\{\ssum\limits_{i\neq
k}\w{a}_iC_{ik}\}\vspace*{-1mm}\]

We apply inequality (\ref{7}) to the second term. We use
that $d\nu =g\, d\mu$, so $d\nu_k
=\ffrac{g}{\mu_k(g)}d\mu_k$ and obtain
\vspace*{-1mm}\[\vv \nu_y\{\int f\, d\mu_k(\cdot\vv y)-\int f\, d\nu_k (\cdot\vv
y)\}\vv =\vspace*{-1mm}\]
\vspace*{-1mm}\[=\vv\nu_y\{ \int [f-\mu_k(f)](d\mu_k-\frac{g}{\mu_k(g)}d\mu_k )\}\vv
=\vspace*{-1mm}\]
\vspace*{-1mm}\[=\vv \mu_y\{\frac{g}{\mu_k(g)}\int (f-\mu_k(f))(g-\mu_k(g))d\mu_k\}\vv
\vspace*{-1mm}\]

The result of integration on $\RR_k$ doesn't depend on variable
$x_k\in\RR_k$, therefore we continue
\vspace*{-1mm}\[\vv \mu_y\{\frac{g}{\mu_k(g)}\int (f-\mu_k(f))(g-\mu_k(g))d\mu_k\}\vv
=\vspace*{-1mm}\]
\vspace*{-1mm}\[=\vv \mu \{ \int_{\RR_k}(f-\mu_k(f))(g-\mu_k(g))d\mu_k\}\vv\leq\vspace*{-1mm}\]
\vspace*{-1mm}\[\leq \sup \,
cov^{1/2}_{\mu_k}(f,f)cov^{1/2}_{\mu_k}(g,g)\leq\de_k(f)\de_k(g)=
b_k\de_k(f)\vspace*{-1mm}\]

Finally we have obtained the estimate on (\ref{10})
\vspace*{-2mm}\begin{equation}
\vv\mu (f)-\nu (f)\vv\leq
\ssum\limits_{j\neq k}\w{a}_j \de_j (f)+\de_k(f)\{\ssum\limits_{i\neq
k}\w{a}_i\la C_{ik}+b_k\}
\Label{kko}\vspace*{-2mm}\end{equation}

By iteration of (\ref{kko}) like in \cite{Fo,GrD,Ku}
 one achieves that $(\w{a}\la
C+b)$ is an estimate too
\vspace*{-2mm}\begin{equation}
\vv\mu (f)-\nu (f)\vv\leq\ssum\limits_{k\in\zd}\{ \w{a}_i\la
C_{ik}+b_k\}\de_k(f)
\Label{f6}\vspace*{-2mm}\end{equation}

The vector
$b\ssum\limits_{n=0}^\infty (\la C)^n$ is also an
estimate because of the following convergence
in $\ell_\infty (\zd )$
\vspace*{-1mm}\[ b\ssum\limits_{n=0}^N(\la C)^n+\w{a}(\la C)^{N+1}
\to b\ssum\limits_{n=0}^\infty (\la C)^n,\
\ \ \ N\to\infty\vspace*{-1mm}\]

Thus we achieve estimate \cite{Fo,GrD,Ku}
\begin{equation}
\vv cov_\mu (f,g)\vv =\vv \int\limits_{\RR^\zd}f\, d\nu
-\int\limits_{\RR^\zd}f\, d\mu\vv
\leq \ssum\limits_{k,j\in\zd}D_{kj}\de_k(f)\de_j(g)
\Label{f7}\end{equation}
for $D=\ssum\limits_{n=0}^\infty (\la C)^n$. Therefore
\vspace*{-1mm}\[\vv cov_\mu (f,\tau_i g)\vv\ee^{d(i,0)}
\leq\ssum\limits_{k,j\in\zd}\ee^{d(j,k)}D_{jk}\ee^{d(k,0)}\de_k(f)
\ee^{d(i,j)}\de_{j-i}(g)\vspace*{-1mm}\]
Summing up on $i\in\zd$ we have the required decay of correlations for
$g>0$.

The case of arbitrary $g\in C^1_{b,cyl}(\RR^\zd )$ with
$\ssum\limits_{k\in\zd}\ee^{d(k,0)}\de_k(g)<\infty$ is obvious due to the
identity $cov_\mu (f,c_1g+c_2)=c_1\, cov_\mu (f,g)$ \ \ \ $\BBox$

\vspace{4mm}

\think{Theorem}{2.}{
Under conditions A--C the set of Gibbs measures $\gml$ with
condition
\vspace*{-2mm}\begin{equation}
m_\mu =\ssup\limits_{k\in\zd}\int\limits_{\RR^\zd}\rho^2(x_k,0)d\mu(x) <\infty
\Label{a}\vspace*{-2mm}\end{equation}
is nonempty.

Moreover, at the coupling interaction constant $\la\in [0,1/\ga_d)$,
the Gibbs measure $\w{\mu}$ of Theorem 1 fulfills estimate
\begin{equation}
\ssup\limits_{k\in\zd}\int\limits_{\RR^\zd}\exp \{ ax_k^2\}
d\w{\mu}\leq
\exp (\ffrac{a}{\vep -2a})
\Label{5}\end{equation}
for all $a\in [0,\vep /2)$.
}

\noindent{\it Proof.}\ \ \ Let
\vspace*{-1mm}\[{\cal U}_\Lambda =\ssum\limits_{k\in\Lambda}F(x_k) +\la\ssum\limits_{\{
k,j\}\subset\Lambda}G_{k-j}(x_k-x_j)\vspace*{-1mm}\]
and consider the family of Gibbs measures $\{\mu_\Lambda\}$ with the
free boundary conditions in the finite volumes $\Lambda\subset\zd$
\vspace*{-1mm}\[d\mu^0_\Lambda =\ffrac{1}{Z}\ee^{-{\cal U}_\Lambda}dx_\Lambda\vspace*{-1mm}\]

The potentials $({\cal U}_\Lambda )^{\pr\pr}\geq \vep I$ are convex, so
the measures $\mu^0_\Lambda$ satisfy inequality (\ref{fff}) in form
\cite{BaEm,DaSi}
\vspace*{-1mm}\[cov_{\mu^0_\Lambda}(f,f)\leq\frac{1}{\vep}\int\limits_{\RR^\Lambda}
\ssum\limits_{k\in\Lambda}\vv\dd_k f\vv^2 d\mu^0_\Lambda\vspace*{-1mm}\]

Substituting $f=x_k$ and using that $\int_{\RR^\Lambda}x_k \,
d\mu^0_\Lambda \equiv 0$ by the symmetry of $\mu^0_\Lambda$ we have that
uniformly on $\Lambda$ and $k\in\Lambda$
\vspace*{-2mm}\begin{equation}
\ssup\limits_{\Lambda\subset\zd,\
k\in\Lambda}\int\limits_{\RR^\Lambda}x_k^2 \ d\mu^0_\Lambda \leq 1/\vep
\Label{4}\vspace*{-2mm}\end{equation}

The convexness of the potentials ${\cal U}_\Lambda$ also imply the
Log-Sobolev inequality for the measures $\{
\mu^0_\Lambda\}$ \cite{BaEm}
\vspace*{-2mm}\begin{equation}
\int_{\RR^\Lambda}f^2\ln f^2 \, d\mu^0_\Lambda -\int_{\RR^\Lambda}
f^2\, d\mu^0_\Lambda \ln\int_{\RR^\Lambda}f^2\,
d\mu^0_\Lambda\leq\frac{2}{\vep}\int_{\RR^\Lambda}
\ssum\limits_{k\in\Lambda} \vv\dd_k f(x_\Lambda )\vv^2
d\mu^0_\Lambda (x_\Lambda )
\Label{b}\vspace*{-2mm}\end{equation}

Fix $\Lambda\subset\zd$ and $k\in\Lambda$.
Consider increasing on $n\geq 1$ sequence of functions
\vspace*{-1mm}\[f_{n}=\left\{
\begin{array}{l}
-n,\ \ \ x_k<-n\\
x_k,\ \ \ \vv x_k\vv\leq n\\
n,\ \ \ x_k>n
\end{array}\right.\vspace*{-1mm}\]

Like in \cite{DaSi} introduce
sequence of functions $h_{n}(a)=\int\limits_{\RR^\Lambda}
\exp (af_{n}^{2})d\mu^0_\Lambda\geq 1$ on half-line $a\in [0,\infty )$,
increasing on both $a$ and $n$ with all derivatives $h_{n}^{(k)}(a)>0$, $a>0$.
Then for $g_{n}=\exp (af_{n}^{2}/2)$
we apply Log-Sobolev \mbox{inequality (\ref{b})}
\vspace*{-1mm}\[ ah_{n}^{\prime}(a)=\int\limits_{\RR^\Lambda}
af_{n}^{2}\exp (af_{n}^{2})d\mu^0_\Lambda
=\int\limits_{\RR^\Lambda} g_{n}^{2}\ln g_{n}^{2}\; d\mu^0_\Lambda\leq\vspace*{-1mm}\]
\vspace*{-1mm}\[\leq\frac{2}{\vep}\int\limits_{\RR^\Lambda}\ssum\limits_{j\in\Lambda}
\vv\dd_j g_n\vv^2 d\mu^0_\Lambda +h_{n}(a)\ln h_{n}(a)\leq\vspace*{-1mm}\]
\vspace*{-1mm}\[\leq \frac{2}{\vep}a^{2}\int\limits_{\RR^\Lambda}
f_{n}^{2}\exp (af_{n}^{2})d\mu^0_\Lambda +h_{n}(a)\ln h_{n}(a)\vspace*{-1mm}\]
Therefore the family $h_{n}(a)$, increasing on both $n$ and $a\geq 0$,
$h_{n}(0)=1$, satisfy inequality
$a(1-\frac{2a}{\vep})h_{n}^{\prime}(a)\leq h_{n}(a)\ln h_{n}(a)$.
To find the major function we must set $h(0)=1$ and take the highest
growth of its derivative, so
$a(1-\frac{2a}{\vep})h^{\prime}(a)= h(a)\ln h(a)$
and $h(a)=\exp (\displaystyle{\frac{aD}{1-{2a}/{\vep}}})$
for some $D$. The restriction on $D$ we obtain from the highest growth
of $h_{n}$ at zero
\vspace*{-1mm}\[ h_{n}^{\prime}(0)=\int\limits_{\RR^\Lambda}
f_{n}^{2}d\mu^0_\Lambda\leq\int\limits_{\RR^\Lambda}
x^2_k d\mu^0_\Lambda (x) =D<\infty\vspace*{-1mm}\]
and achieve estimate $h_{n}(a)\leq \exp \{ \displaystyle{\frac{a}{1-{2a}/{\vep}}}
\int_{\RR^\Lambda}
x_k^2 d\mu^0_\Lambda (x)\}$. Tending $n\to\infty $ we obtain
the estimate of the next form at
$a\in [0,\vep /2)$
\vspace*{-1mm}\[\int_{\RR^\Lambda}\exp (ax_k^2)d\mu^0_\Lambda \leq \exp
(\ffrac{a}{1-2a/\vep }\int_{\RR^\Lambda}x_k^2 d\mu_\Lambda^0 )\vspace*{-1mm}\]
which by (\ref{4}) gives
\vspace*{-2mm}\begin{equation}
\all a\in [0,\vep /2)\ \ \ \ \ssup\limits_{\Lambda\subset\zd ,\ k\in\Lambda}\int_{\RR^\Lambda}\exp
(ax_k^2)d\mu_\Lambda^0 <\exp (\ffrac{a}{\vep -2a})
\Label{100}\vspace*{-2mm}\end{equation}

Compactness of the function $\exp (ax_k^2)$
leads by the Prochorov's theorem \cite{Sin}
to the existence of the weak local
limit $\w{\mu}$
\vspace*{-1mm}\[\llim\limits_{\Lambda\nearrow\zd}\int_{\RR^\Lambda} f(x_\Lambda
)d\mu^0_\Lambda =\int_{\RR^\zd}f(x)d\w{\mu}\vspace*{-1mm}\]
on any cylinder function $f\in C_{b,cyl}(\RR^\zd )$.

Due to the finiteness of the interaction radius B the
limit measure $\w{\mu}$ has the conditional measures $\{\mu_\Lambda\}$
in the finite volumes, i.e. $\w{\mu}\in \gml$ and the set of the Gibbs
measures is non-empty.
From (\ref{100}) we also have that the measure $\w{\mu}$ is tempered, i.e.
\vspace*{-1mm}\[\ssup\limits_{k\in\zd}\int_{\RR^\zd}\exp (ax_k^2)d\w{\mu} <
\exp (\ffrac{a}{\vep -2a}),\ \ \ \ a\in [0,\vep /2)\vspace*{-1mm}\]
which obviously gives the statement (\ref{5}).\ \ \ $\BBox$

\vspace{4mm}
\noindent {\bf Model 1.}\ \ \ Let the potentials be defined by
\vspace*{-1mm}\[F(x_k)=(1+x_{k}^{2})^{2n+1}\ \ \ \ \&\ \ \ \ G_{k-j}(x_k-x_j)
=b_{k-j}(x_k-x_j)^{2n+2}\vspace*{-1mm}\]
and assume that the coefficients $\{ b_j\}_{j\in\zd}$ satisfy
\vspace*{-1mm}\[\forall j\in\zd\ \  b_j\geq 0\ \ \& \ \
\exists r_{0} \ \forall \vert j\vert >r_{0}: b_j=0\vspace*{-1mm}\]

Then for
\vspace*{-1mm}\[ 0\leq \lambda <\displaystyle{\frac{1}{(n+1)2^{2n+1}\Vert
b\Vert_{d}}},\ \ \ \ \ \Vert b\Vert_{d}=\ssum\limits_{j\in\zd}b_j
e^{d(j,0)} <\infty\vspace*{-1mm}\]
the statements of Theorems 1,2 are valid.

\vspace{4mm}

\noindent{\bf Model 2.}\ \ \ \begin{minipage}[t]{5in}
\it Lattice spin system on Riemannian manifold.
\end{minipage}

Denote $M=M_k,\ k\in\zd$ a noncompact Riemannian manifold with covariant
derivative $\dd_k$ and Ricci curvature tensor $Ric_k$.

Let potentials $F_k(x_k),\ G_{kj}(x_k,x_j)$ satisfy

1) \ $F_k\in C^2(M),\ \is\vep >0\ \all x_k\in M_k\ \ Ric_k +\dd_k\dd_k
F(x_k)\geq\vep$

2) \ $G_{kj}\in C^2(M\times M),\ \is\al\in\RR^1\ \
\dd_k\dd_kG_{kj}(x_k,x_j)\geq -\al ,\ \ k,j\in\zd$

\mbox{ }\ \ \ and $G_{kj}\equiv 0$, for $\vv k-j\vv\geq r_0$.

3) $\al_{k,j}=\ssup\limits_{x\in M^\zd}\vvv
B^{-1/2}(x_k)B^{-1/2}(x_j)\dd_k\dd_j G_{kj}(x_k,x_j)\vvv_{TM_k\times
TM_j}<\infty$

\noindent where $B(x_k)=Ric_k(x_k)+\dd_k\dd_k F_k(x_k)$ and
$\vvv\cdot\vvv_{TM_k\times TM_j}$ is a standard Hilbert norm on tangent
space to $M_k\times M_j$.

Then for $\la\in [0,\min (\vep /\al (2r_0+1)^d , 1/\ga_d))$ the lattice
system, described by Hamiltonian
\vspace*{-1mm}\[
H=\ssum\limits_{k\in\zd}F_k(x_k) +\la \ssum\limits_{\vv k-j\vv\leq
r_0}G_{kj}(x_k,x_j)
\vspace*{-1mm}\]
has exponentially fast decay of correlations and Gibbs measure is
unique \cite{AAW}. Above $\ga_d =\ssup\limits_{k\in\zd}\ssum\limits_{j\in\zd}
\ee^{d(k,j)}\al_{kj}$.

This result is achieved by the scheme of this paper. One
needs to consider
\vspace*{-1mm}\[
\de_k(f)=\ssup\limits_{x\in M^\zd} \vvv (Ric_k+\dd_k\dd_k
F(x_k))^{-1/2}\dd_k f(x)\vvv_{TM_k}
\vspace*{-1mm}\]
and apply in corresponding places the following generalization of
Brascamp-Lieb inequality (\ref{7}) to the case of arbitrary
Riemannian manifold \cite{AAW}:  $\;$under condition $\is \vep >0\
Ric+\dd\dd F\geq \vep$ we have
\vspace*{-1mm}\begin{equation}
cov_\mu (f,f)\leq \int_M <(Ric +\dd\dd F)^{-1}\dd f,\dd f> d\mu ,\ \ \
f\in C_b^1 (M)
\vspace*{-1mm}\Label{fff}\end{equation}
with probability measure $d\mu =\ee^{-F}d\sigma$ ($\sigma$ denotes
Riemannian volume on manifold $M$) and
Riemannian pairing $<\cdot ,\cdot >$
on tangent space to manifold.

Developing idea of Helffer \cite{Hel,NaSp} we can shortly explain
inequality (\ref{fff}) next way.
Take $u=\dd H_\mu^{-1}(g-\int g\, d\mu )$ for
$H_\mu =\dd^*_\mu\dd $ with dual gradient $\dd^*_\mu v=-div\, v+
<\dd F,v>$, then $\dd^*_\mu u=H_\mu H_\mu^{-1}(g-\int g\, d\mu )=g-\int g\,
d\mu$ and we have
\vspace*{-1mm}\[
cov_\mu (g,g)=\int g(g-\int g \, d\mu )d\mu =\int <u,\dd g> d\mu =
\vspace*{-1mm}\]\vspace*{-1mm}\[
=\int <(H_\mu +Ric+\dd\dd F)^{-1}\dd g,\dd g> d\mu \leq \int_M <(Ric+\dd
\dd F)^{-1}\dd g,\dd g>d\mu
\vspace*{-1mm}\]
where we used positivity of $H_\mu$ and $u=(H_\mu +Ric +\dd\dd
F)^{-1}\dd g$ by a simple commutation $\dd g=\dd\dd^*_\mu u=(\dd^*_\mu\dd
+Ric+\dd \dd F)u$.

\vspace{3mm}
\noindent {\bf Acknowledgements.}\ \ \ We wish to thank sincerely Prof.
S.Albeverio for the attention to the paper, warm hospitality in
the BiBoS Research Center and great help to make \cite{BePi,Pic}
available for us. We would like to express our appreciation
to Prof. L.Gross for the invaluable recommendations on \cite{AAW}
and Prof. T.Spencer for the prompt contact
with new information on $(\nabla\varphi )^4$ lattice models.

\vspace{5mm}

\normalsize
\bibliographystyle{plain}

\end{document}